\documentclass{article}
\usepackage{arxiv}

\pdfoutput=1 

\usepackage[utf8]{inputenc} 
\usepackage[T1]{fontenc}    
\usepackage{hyperref}       
\usepackage{url}            
\usepackage{booktabs}       
\usepackage{amsfonts}       
\usepackage{nicefrac}       
\usepackage{microtype}      
\usepackage{lipsum}

\usepackage{amsmath,amssymb}
\usepackage{graphics,graphicx}
\graphicspath{{./Figures/}}
\usepackage{color}
\usepackage{threeparttable}
\usepackage{url}

\DeclareMathOperator*{\argmin}{arg\, min}

\begin{document}

\title{A Many Objective Optimization Approach for Transfer Learning in EEG Classification}

\author{
  Monalisa Pal\\
  Machine Intelligence Unit	 \\
  Indian Statistical Institute\\
  Kolkata - 700108, India\\
  \texttt{monalisap90@gmail.com} \\
   \And
 Sanghamitra Bandyopadhyay\\
  Machine Intelligence Unit	 \\
  Indian Statistical Institute\\
  Kolkata - 700108, India\\
  \texttt{sanghami@isical.ac.in} \\
     \And
 Saugat Bhattacharyya\\
  BCI-NE Lab, School of Computer Science \& Electronics Engineering	 \\
  University of Essex\\
  Colchester CO4 3SQ, United Kingdom\\
  \texttt{saugatbhattacharyya@live.com} \\
}

\maketitle

\begin{abstract}
In Brain-Computer Interfacing (BCI), due to inter-subject non-stationarities of electroencephalogram (EEG), classifiers are trained and tested using EEG from the same subject. When physical disabilities bottleneck the natural modality of performing a task, acquisition of ample training data is difficult which practically obstructs classifier training. Previous works have tackled this problem by generalizing the feature space amongst multiple subjects including the test subject. This work aims at knowledge transfer to classify EEG of the target subject using a classifier trained with the EEG of another unit source subject. A many-objective optimization framework is proposed where optimal weights are obtained for projecting features in another dimension such that single source-trained target EEG classification performance is maximized with the modified features. To validate the approach, motor imagery tasks from the BCI Competition III Dataset IVa are classified using power spectral density based features and linear support vector machine. Several performance metrics, improvement in accuracy, sensitivity to the dimension of the projected space, assess the efficacy of the proposed approach. Addressing single-source training promotes independent living of differently-abled individuals by reducing assistance from others. The proposed approach eliminates the requirement of EEG from multiple source subjects and is applicable to any existing feature extractors and classifiers. Source code is available at \url{http://worksupplements.droppages.com/tlbci.html}.
\end{abstract}


\section{Introduction}
Brain-Computer Interfacing (BCI) bypasses the biological neuro-muscular pathway to accomplish a task and creates a thought-based electronic communication channel between the idea-generating unit, i.e., brain, and the idea-implementing unit, i.e., an external device with an intermediate computing unit to process ideas and generate control signals for driving the external device \cite{MBEC14,BCIStateofArt1}. The action of the external device provides a feedback to the subject to adjust his neural network for generating corrective brain signals until the desired task is accomplished \cite{BSPC15}. For collecting brain signals \cite{BCIStateofArt1,BCIStateofArt2,ASOC4}, there are several methods each having its own advantages and disadvantages. Some of these methods are electroencephalography (EEG), magnetoencephalography (MEG), functional magnetic resonance imaging (fMRI), electrocorticography (ECoG) and function near-infrared spectroscopy (fNIRS). However, EEG has been the choice of brain signal acquisition medium for several of its merits like cost-effectiveness, high temporal resolution, non-invasiveness and convenience in terms of portability\cite{MBEC14, BCIStateofArt1,BSPC15,MicroComBCI16,ASOC1,ASOC2,ASOC3}. Such BCI systems have a plethora of applications like virtual gaming, navigation, communication, rehabilitation, robotics, health-care, military services and many more \cite{Appl2,HMS3,HMS4,rehabscene1}. 

In several of these applications, recognition of motor imagery tasks \cite{MicroComBCI16,conecct2014,semcco13,spcom14,HMS1} from EEG signals is a preliminary step which obeys supervised learning techniques \cite{TheodoridisCh10}. The steps involved in such a typical EEG-based BCI system \cite{BCIStateofArt1,BCIStateofArt2} are as follows:
\begin{enumerate}
	\item EEG signal acquisition from subjects whose mental states are to be decoded and denoising the raw signals
	\item Processing the signals for feature extraction
	\item Classification of the mental states using a pre-trained classifier
	\item Generating control signals based on predicted brain signal in order to operate an external device
\end{enumerate}

Literature is galore with examples of EEG-based motor imagery signal classification. Distinguishable features and efficient classifiers for this purpose has been analysed in numerous studies, as in \cite{lit2,lit3,lit7,conecct2014,lit5,lit6}. In order to improve accuracy and to reduce processing time, feature reduction has been explored where the dimension of the reduced feature space is either an user-input \cite{MicroComBCI16,conecct2014,semcco13,featsel2013} or automatically determined \cite{MBEC14,spcom14}. However, in this traditional BCI framework, training of the classifier becomes challenging as EEG signal from different subjects \cite{jour36,R3_6_22,proc15,R3_6_46} or even from different sessions \cite{BSPC15,lit8,SNRE1} of the same subject is deemed to be unsuitable due to high variance of EEG signals from time to time and from person to person for the same task \cite{intersubjEEG1,intersubjEEG2}.

To avoid repeated training of classifiers, the researchers have developed effective means to improve inter-session EEG classification like using interval type-2 fuzzy classifier \cite{BSPC15}, spectral kernel discriminant analysis \cite{lit8}, stacked regularized linear discriminant analysis \cite{SNRE1}, and so on, whereas, research in inter-subject EEG classification is relatively recent and has several open challenges.

Rehabilitative BCI assists people who are able to think but unable (or disabled) to perform mobility-related task \cite{BSPC15,rehabscene1,MicroComBCI16,HMS1}. Collecting large volume of training data is difficult in such situations. It is also infeasible to imagine all possible kinds of tasks and thus, impractical to have pre-trained classifiers for an exhaustive list of tasks. Hence, recent researches are focusing on inter-subject EEG classification. 

Reviews on various transfer learning approaches, proposed for BCI problems, are presented in \cite{Review1,Review2}. Broadly, there are three groups of transfer learning approaches. The first group is called the feature representation transfer group where improved features are presented in order to encode the common BCI knowledge across multiple subjects \cite{jour36,proc15,jour35,jour34}. The second group is called the instance transfer group in which learning occurs based on parts of data where the distribution is observed to be similar \cite{R3_6_22, R3_6_55}. The third group is called the classifier transfer group which deals with domain adaptation methods \cite{R3_6_46,R3_6_1,R3_6_39,newlit1} that are capable of handling different data distributions. These inter-subject transfer methods are dependent on multiple source subjects which are subjected to availability. Hence, either a large volume of data has to be stored (increased memory requirement) or has to be electronically transmitted from a remote location (increased time requirement) when needed. Storing EEG data for all tasks is again impractical, mostly for stand-alone systems with restricted memory. Also inter-subject EEG features which are based on common spatial patterns (CSP) cannot be directly used for multi-class classification. The proposed work is first of its kind to address an approach using single source subject and is invariant to the features, classifier and the number of tasks.

In this work, right hand/foot motor imagery EEG signals are classified from the BCI Competition III Data set IVa where linear Support Vector Machine (SVM) is trained by the weighted features of one subject (source subject) but decodes the intentions for an unrelated subject (target subject) using power spectral density estimates as base features. A many-objective optimization (MaOO) model of transfer learning is proposed which yields a weight matrix by maximizing source-trained target EEG classification performance. The weight matrix encodes information about the source-target EEG association. To the best of the authors' knowledge, such an attempt to address the issue of inter-subject EEG classification using a MaOO approach, where EEG data from only one source subject is required during training, is first of its kind and here lies the novelty of the work.

Rest of the paper is organized as follows. Section \ref{PS} presents the related literature survey, their drawbacks and the proposed framework of the inter-subject EEG classification problem to overcome the drawbacks. The theoretical aspects (in Section \ref{Theory}) and the experimental specifications (in Section \ref{Exp}) of the proposed many-objective optimization model for the transfer learning problem are outlined next. The efficacy of the proposed approach is undermined by the results presented in Section \ref{RnD} and the chief observations are discussed in Section \ref{Discuss}. Finally, the paper is concluded in Section \ref{Con} with some future directions.

\section{Motivation behind the proposed work}
\label{PS}
A major application of BCI is in rehabilitation where the natural modality of performing a task is disabled for the subject and hence an artificial channel is established for communication between the mobility-disabled subject and prosthetic devices \cite{Appl1}. A large amount of training data for several tasks cannot be gathered from the mobility-disabled individual. In this or similar scenarios, recognizing the subject\textquoteright s (disabled) mental states (EEG) using classifier trained by another subject\textquoteright s (healthy) EEG is essential \cite{rehabscene1,rehabscene2}. This motivates the researchers to develop efficient inter-subject motor imagery EEG classification schemes. In this section, the existing approaches to address the inter-subject EEG classification and their limitations are discussed. Following this, the distinctive characteristics of the proposed approach are highlighted.

\subsection{Limitations of previous works}
 \label{lit}
Due to inter-session and inter-subject variability, transfer learning has been in limelight in the recent years in the BCI community. A few recent researches and their limitations are mentioned here.
\begin{enumerate}
	\item The studies in \cite{R3_6_22,R3_6_1,newlit1,newlit5} deal with session-to-session transfer. Experiments using these frameworks to validate their applicability in inter-subject transfer are yet to be done. 
	
	\item The approach used in \cite{R3_6} is very similar to the proposed approach where active learning is used to predict the labels of unlabelled data using labelled data. Here, variability between labelled and unlabelled EEG data is introduced due to the use of separate EEG headsets during signal acquisition. However, the approach claims that in order to improve classification performance, same number of labelled samples from the target set are required \cite{R3_6}. Our proposed work is free from any such constraint. Moreover, the experiment \cite{R3_6} has been performed using the data from the same subject and hence, does not demonstrate its compatibility with inter-subject transfer. 
	
	\item The approaches in \cite{jour36,R3_6_22,proc15,R3_6_46,jour35,R3_6_1,newlit5} use CSP or CSP with modified parameters, as features. However, CSP is applicable to binary classification only. Thus, direct extension of these methods for multi-class classification is not possible. All combinations of one versus one or one versus all learning has to be analysed for multi-class classification which is computationally costly. In contrast to these works, the framework of the proposed approach is independent of the features used and hence, is directly extensible to multi-class classification.
	
	\item The works in \cite{jour36,R3_6_22,jour35,jour34,R3_6_55} report the performance of their approaches only in terms of classification accuracy and/or error rate. Based only on these results, it is difficult to assess the efficacy of these approaches for problems where class imbalance occurs. In our proposed work, as more number of performance metrics are reported, the claim of a better and a robust performance of the framework holds more strongly.
	
	\item The work in \cite{R3_6_1} deals in modifying Extreme Learning Machine (ELM) to make it adaptive such that it can classify data from evaluation session when trained by data from calibration session. Similarly, the works in \cite{R3_6_22,proc15,R3_6_46,jour35,newlit5} uses Linear Discriminant Analysis (LDA) as a base classifier and modifies the covariance matrix or uses covariate shift adaption to make the framework suitable for transfer learning. Another popular classifier viz. $k$-Nearest Neighbors ($k$NN) has also been tailored to perform inter-subject transfer learning \cite{R3_6_55}. The framework proposed in this manuscript has advantage over these works, as it is applicable to any supervised learning algorithm.
	
	\item The works in \cite{jour36,proc15,jour35,jour34,R3_6_55,newlit2,newlit4,newlit6} proposes different approaches to perform inter-subject transfer. However, each of these approaches involve learning from multiple source subjects to predict the intentions of the target subject. In the proposed work, classification of target subject's EEG is achieved by learning from only one source subject. This is hugely advantageous over the existing approaches as data acquisition from multiple healthy source subjects is time-consuming and is constrained by the availability of the subjects.
\end{enumerate}
\subsubsection{Characteristic features of the proposed work}
 \label{merits}
The proposed framework presents a method addressing inter-subject EEG classification which not only uses a single source subject to obtain the training data but also can directly be used with any existing multi-class features and classifiers, without any modification, for prediction of two or more thought-based tasks. This saves computational resources and time, both of which are important for a real-time stand-alone system. All the characteristic features of the proposed work are enlisted as follows:
\begin{enumerate}
	\item It is targeted to predict unlabelled data using a trained classifier where due to high variability between labelled and unlabelled data, traditional classifiers perform poorly.
	
	\item It uses labelled data only from one another source and thereby, tries to reduce the dependency on external sources of data.
	
	\item It is modelled in such a fashion that any existing technique of feature extraction and classification could be used with this framework. This is beneficial because:
	
	\begin{itemize}
		\item the framework could be directly extended to be used for multi-class classification with appropriate features and classifiers, and
		\item features/classifiers designed for inter-session EEG classification \cite{BSPC15,lit8,SNRE1} could be integrated in this framework to address both inter-subject and inter-session EEG classification simultaneously.
	\end{itemize}

	\item While regularising the supervised learning framework, it inherently enhances several classification performance metrics such that the issue of class imbalance could also be taken into account.
\end{enumerate}

As per the author's awareness, a method involving all these features to perform transfer learning for BCI is not available in the BCI literature and here lies the novelty of the proposed work.

\section{Theoretical concept behind the proposed work}
 \label{Theory}
Due to the huge variance in inter-subject EEG signal \cite{jour36,proc15,jour35,jour34}, it can be said that the source and target domains (i.e. the distribution of source and target EEG signals given their intentions, $P(S^{subj}|y^{subj})$ with $subj\in\{src,tar\}$) are different, but as both are of same kind (i.e. both $S^{src}$ and $S^{tar}$ are motor imagery EEG signals), they are related. On the other hand, the source and target tasks (i.e. intentions $y^{src}$ and $y^{tar}$ of source and target subjects) are the same. This is formally expressed by Eq. \eqref{eqn: TTL}. In this application, only a part of the target domain data is required during training. Such a scenario becomes a typical case of transductive transfer learning \cite{TLsurvey}.
\begin{equation}
	\begin{split}
		&\text{Source data and labels:} \{(S^{src}, y^{src})\}\\
		&\text{Target data and labels:} \{(S^{tar}, y^{tar})\}\\
		&\text{where, } P(S^{src}|y^{src})\neq P(S^{tar}|y^{tar}) \text{ but } y^{src} = y^{tar}
	\end{split}
	\label{eqn: TTL}
\end{equation}

\begin{figure}[ht]
	\centering
	\includegraphics[width = 0.5\linewidth, bb = 80 50 420 328, clip = true]{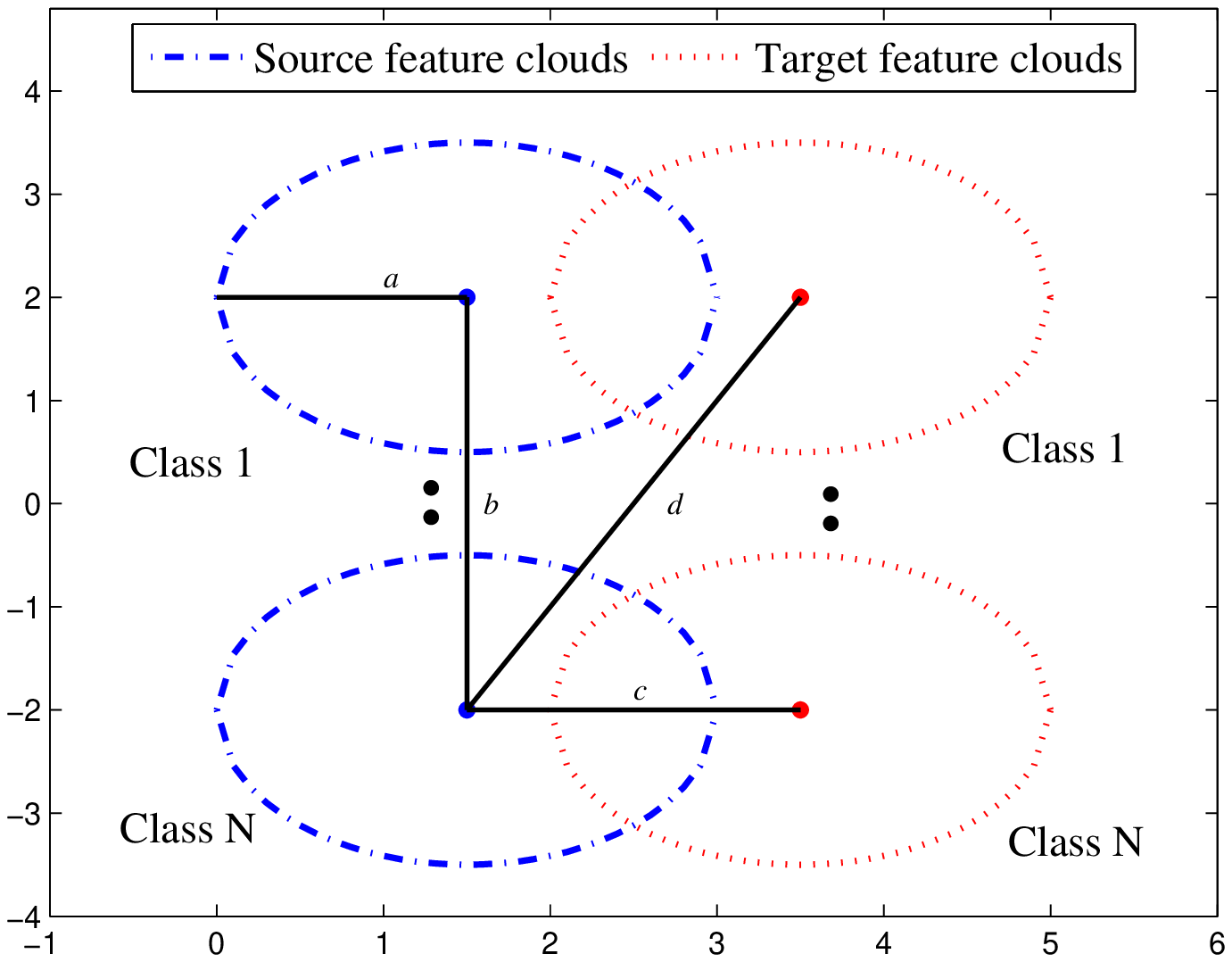}
	\caption{Illustration of Feature Space to Demonstrate the Optimization Goals.}
	\label{fig:scene}
\end{figure}

The idea to tackle this problem is as follows. As the features of same class of EEG data from same subject has less variance, the feature vectors tend to form a compact cloud (a bounded entity in the feature space). Such clouds are well-separated for different classes (or intentions) of a single subject in a single session. Unlike a typical pattern recognition system \cite{TheodoridisCh10}, such feature clouds of different subjects for the same class of EEG data are partially or fully non-overlapping (Fig. \ref{fig:scene}). To facilitate classification, this overlap should be increased. This requires transformation of the features which can be achieved by linearly projecting the features to another space. The weights for linear projection should improve discrimination of features such that:
\begin{itemize}
	\item intra-subject intra-class feature clouds remain compact (by minimizing distances marked $a$ in Fig. \ref{fig:scene}),
	\item intra-subject inter-class feature clouds are well-separated (by maximizing distances marked $b$ in Fig. \ref{fig:scene}),
	\item inter-subject intra-class feature clouds have significant overlap (by minimizing distances marked $c$ in Fig. \ref{fig:scene}),
	\item inter-subject inter-class feature clouds are well-separated (by maximizing distances $d$ in Fig. \ref{fig:scene}).
\end{itemize} 

To describe linear projection in terms of the BCI application, let $J^{subj}$ be a feature matrix which is extracted from the EEG signal i.e. $J^{subj}$ is a function of $P(S^{subj}|y^{subj})$. It has $N$ rows (instances) of $D$-dimensional feature vectors. Let $W$ be a weight matrix of order $d\times D$ where $d$ is the dimension of the projected space. The projected feature matrix, $J_W$, is generated by Eq. \eqref{eqn: Proj}.  
\begin{equation}
	\begin{split}
		& [J^{subj}]_{N\times D}[W^T]_{D\times d}=[J_W^{subj}]_{N\times d}\\
		& \text{where, } subj\in\{src,tar\}
	\end{split}
\label{eqn: Proj}
\end{equation}

The four characteristics of linear projection help to improve inter-subject EEG classification performance. Hence, $W$ for linear projection can be generated by maximizing source-trained target EEG classification performance. Only a small volume of target EEG should be used while designing the $W$ to avoid the cognitive fatigue problem which is as follows. Repetition of thought-based tasks to provide large training data creates cognitive fatigue \cite{cogload1,HMS2} leading to non-stationarity of the signal which is more enhanced if the target subject is physically disabled \cite{cogload2}. Hence, with the projected features using optimal $W$, the problem defined by Eq. \eqref{eqn: TTL} reduces to Eq. \eqref{eqn: SOL} which can be solved using supervised learning techniques. Thus, this approach can be effectively used for inter-subject EEG classification. 
\begin{equation}
	\begin{split}
		&\text{Projected source features and labels:} \{(J^{src}_W, y^{src})\}\\
		&\text{Projected target features and labels:} \{(J^{tar}_W, y^{tar})\}\\
		&\text{where, } J^{src}_W= J^{tar}_W \text{ but } y^{src} = y^{tar}
	\end{split}
	\label{eqn: SOL}
\end{equation}

\section{Experimental Paradigm}
\label{Exp}
\begin{figure*}[ht]
	\centering
	\includegraphics[width = \linewidth, bb = 30 570 710 780, clip = true]{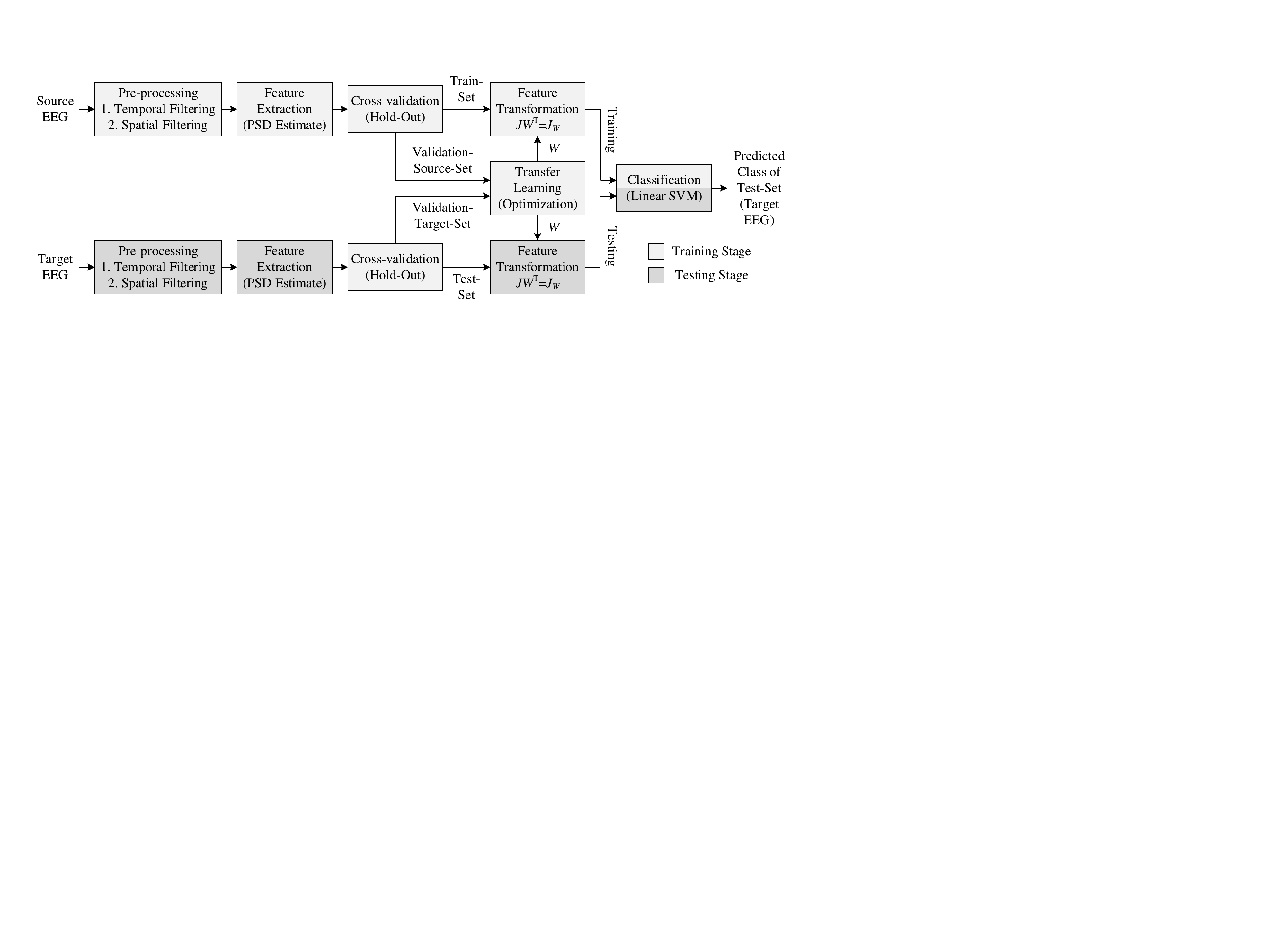}
	\caption{Experimental Framework (phases applicable to testing stage are highlighted in darker shade).}
	\label{fig: framework}
\end{figure*}

BCI system consists of the following stages: data acquisition, pre-processing, feature extraction, cross-validation and classification \cite{TheodoridisCh10}. For the concerned problem, prior to classification, there is an additional step of feature transformation for the proposed transfer learning approach. The entire experimental framework is shown Fig. \ref{fig: framework}.

\subsection{Description of the dataset}
 \label{dataset}
The proposed framework has been tested with a publicly available benchmark dataset – BCI Competition III Dataset IVa \cite{MBEC14,jour36,jour35,jour34,BCIDataset}. 

The specification of the dataset are as follows:
\begin{itemize}
	\item The dataset provides visually stimulated motor imagery EEG data for two classes (Class 1 - right hand and Class 2 - right foot motor imagery).
	\item Out of 128 Ag/AgCl channels of electrode cap (ECI make), 118 channels (modified international 10/20-system placement \cite{placement}) are used.
	\item The experiment for creating the dataset involved five healthy subjects who have subject identifiers as $aa$, $al$, $av$, $aw$ and $ay$, respectively.
	\item The acquired signals, from each subject, are amplified using BrainAmp amplifiers, filtered between 0.05 to 200 Hz, digitized having 16 bit (0.1 $\mu$V) accuracy, and downsampled at 100 Hz. 
	\item From each subject, a total of 280 instances of motor imagery EEG data is randomly acquired for any of the two classes, where the duration of each instance is 3.5 seconds. 
\end{itemize}

Hence, in the dataset, the acquired raw EEG data $S^{subj}$ from each of the subject appear in the form of a matrix where each of the 118 columns represent a time series over the entire experimental period at the sampling rate of 100 Hz and each row represent the signal at an instant acquired by each of the 118 EEG channels.

\subsection{Pre-processing}
The acquired raw EEG data $S^{subj}$ prominently retains the effect of various noises and interference which makes it extremely challenging to extract the full information content from the acquired data by eliminating all the noisy data. In order to reduce a fair share of these unwanted effects, temporal and spatial filtering of the raw EEG data are performed.

\begin{itemize}
	\item \emph{Temporal filtering:} The first step of the proposed approach is to filter the raw EEG to remove environmental and cognitive noises from the signal. Thus, the EEG signals are band-pass filtered at 8 to 25 Hz to contain the $\mu$ band and the central-$\beta$ band EEG for informative motor imagery signals \cite{conecct2014}. This filter is designed using a $6^{\text{th}}$ order elliptical filter with 1dB passband and 50dB stopband ripples \cite{MBEC14,BSPC15}. Elliptical filter is chosen due to its characteristic sharp roll-off and equiripple nature in stopband and passband \cite{MBEC14,BSPC15,conecct2014,spcom14}. 
	
	\item \emph{Spatial filtering:} Further, to reduce inter-channel interference, the EEG signals are spatially filtered using a small Laplacian filter \cite{MBEC14,Dornhegebook}, where the average signals of four nearest neighboring channels is subtracted from the signal of every channel.  
\end{itemize}

It is known from literature that the frontal, central and parietal lobes of the brain \cite{MBEC14,MicroComBCI16,Areas1} are vital for motor imagery signals from a physiological point of view. Henceforth, the data from the seven electrodes, viz., F3, F4, C3, Cz, C4, P3 and P4, which lies over these lobes of interest, are taken into consideration. 

For any novel application, an inter-subject channel selection step needs to be performed independently from the proposed framework, such that the regions of brain excited by neural firing could be identified which are physiologically significant to the concerned application. Further signal processing is concerned with the data from the electrodes covering these identified lobes of interest.

\subsection{Feature Extraction} 
After the preprocessing stage, the EEG data are still in the same format as $S^{subj}$ i.e. the seven selected EEG channels / columns of time-series data corresponding to the entire experimental period. From this data matrix of a subject, the data sub-matrices corresponding to 3.5 seconds of EEG data for each of the 280 instances, are spliced out. Hence, each instance of EEG data is represented as a matrix of order $350 \times 7$ as there are $350$ ($=3.5$ seconds $\times 100$ Hz) sample points for $7$ channels.
 
Next, the characteristic features of the motor imagery EEG signal are determined using power spectral density. Power spectral density (PSD-estimates) is chosen as the feature extraction technique because earlier literature \cite{MBEC14} has demonstrated that PSD has distinguishable properties for this dataset. As EEG signals are non-stationary in nature, PSD is calculated as an average of PSD-estimates of small samples of signals. Overlapping windowed segments help to reduce the variance and give more control on the bias/resolution characteristics (small-segment: more stationary, large segment: more precise PSD approximation) as in Welch\textquoteright s periodogram method \cite{MBEC14,MicroComBCI16} of PSD estimation. 

The specification of PSD estimation technique for feature extraction is as follows:
\begin{itemize}
	\item In this study, a Hamming window of size 50, with 50\% overlap between successive segments, is used to obtain one-sided PSD-estimate of EEG \cite{MBEC14,MicroComBCI16} i.e. Welch\textquoteright s periodogram method \cite{MBEC14,MicroComBCI16} of PSD estimation is used on every EEG instance matrix of order $350 \times 7$.
	
	\item The one-sided PSD-estimate of the 100 Hz EEG signal represents the strength in terms of the logarithm of power content of the signal at integral frequencies between 0 to 50 Hz. 
	
	\item From this, the informative motor imagery signal is extracted. Informative motor imagery signal \cite{conecct2014} is found in $\mu$ band (8-12 Hz) and central-$\beta$ band (16-24 Hz).
	
	\item This yields the power of the EEG signal at 14 integral frequency points for the signal from seven electrodes, viz., F3, F4, C3, Cz, C4, P3 and P4 which are appended channel-wise to yield a ($14\times7=) 98$-dimensional feature vector. This feature vector is obtained for each of the 280 EEG instances. Hence, the feature matrix from source ($J^{src}$) or target ($J^{tar}$) subject is of the order $N\times D=280\times 98$.
\end{itemize}

\begin{figure}[ht]
	\centering
	\includegraphics[width = \linewidth]{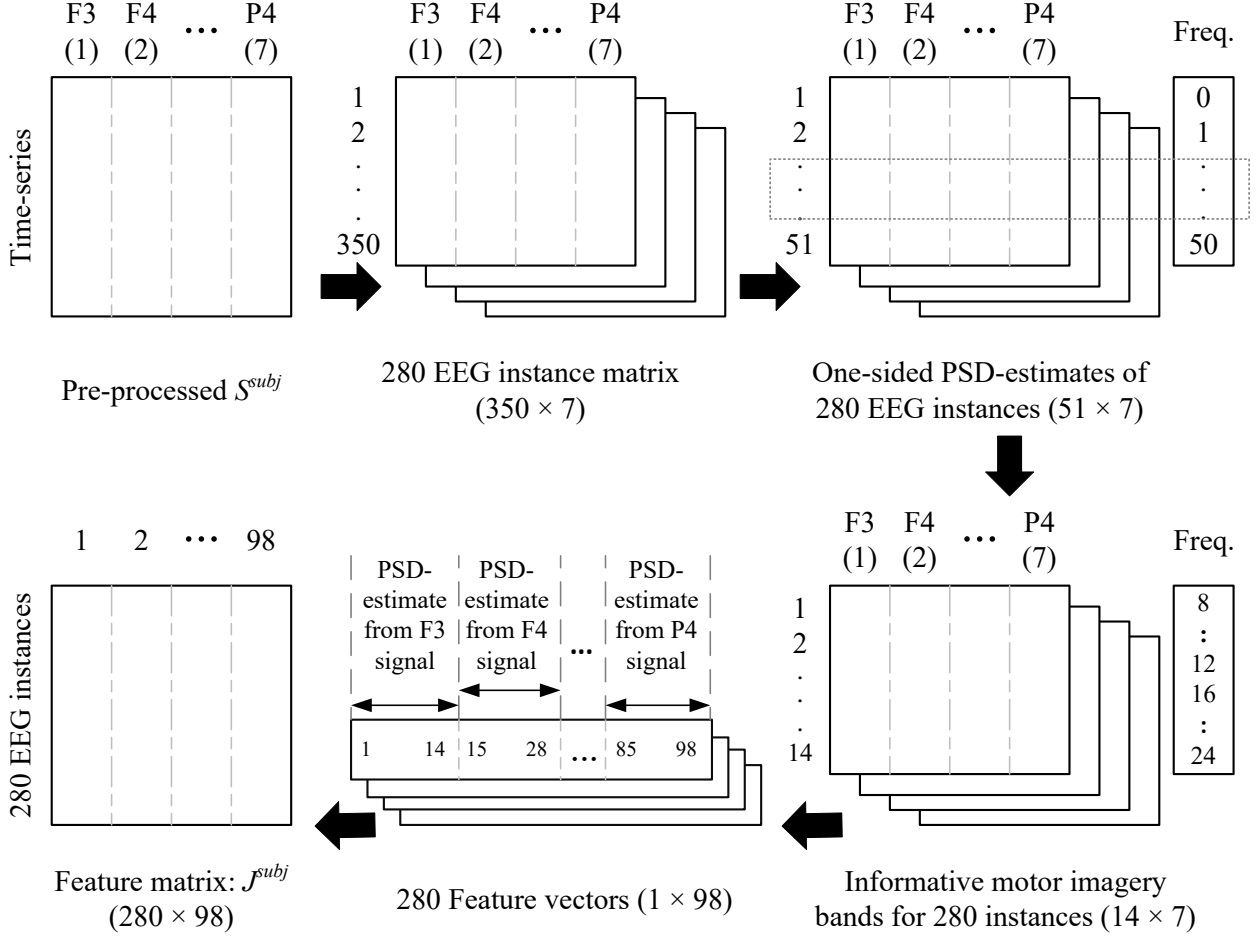}
	\caption{Formation of Feature Vectors from EEG Data.}
	\label{fig:featext}
\end{figure}

These feature extraction steps to obtain $J^{subj}$ from the preprocessed $S^{subj}$ are summarized in Fig. \ref{fig:featext}.

\subsection{Cross-validation}
The choice of number of samples used for different purposes, viz., classifier training (Train-Set), classifier testing (Test-Set), and tuning of classification parameters (Validation-Set), is very crucial for generalization of classifier. In this work, Hold-Out cross-validation \cite{TheodoridisCh10} is used for partitioning each of the source and target datasets into two sets as shown in Table \ref{tab:HoldOut} and Fig. \ref{fig: framework}. The steps of pre-processing, feature extraction, transfer learning (using Validation-Source-Set and Validation-Target-Set), feature transformation (using optimal $W$) and training of classifier (using Train-Set) takes place during the training stage. On the other hand, the steps of pre-processing, feature extraction, feature transformation (using optimized $W$) and testing of classifier (using Test-Set) takes place during the testing stage.
\begin{center}
	\begin{table}[h]
		\caption{Hold-Out Cross-validation Proportions.}
		\label{tab:HoldOut}
		\begin{center}
			\resizebox{0.6\columnwidth}{!}{
			\begin{tabular}{l|l|l}
				\textbf{Data Set} & \textbf{Proportion} & \textbf{Purpose} \\
				\hline
				\hline
				Source & 75\% (randomly & Validation-Source-Set: \\
				(280 samples) & chosen 210 & training classifier used in\\
				& samples) & transfer learning phase\\
				\cline{2-3}
				& 25\% (randomly & Train-Set: \\
				& chosen 70 & final training of classifier\\
				& samples) & with optimal $W$ \\
				\hline
				Target & 25\% (randomly & Validation-Target-Set: \\
				(280 samples) & chosen 70 & testing classifier used in\\
				& samples) & transfer learning phase\\
				\cline{2-3}
				& 75\% (randomly & Test-Set: \\
				& chosen 210 & final testing of classifier\\
				& samples) & with optimal $W$ \\
				\hline
			\end{tabular}	
		}
	\end{center}
\end{table}
\end{center}

\subsection{Transfer Learning using Many-objective Optimization - The Proposed Approach}
\label{TLOP}
The purpose of this stage is to obtain optimal $W$ which maximizes source-trained target EEG classification performance. The proposed approach tries to solve the inter-subject EEG classification problem by training a classifier using the projected source features (source subject\textquoteright s EEG or the Validation-Source-Set) and testing it using a small amount of the projected target features (target subject\textquoteright s EEG or the Validation-Target-Set). The performance of this classification can be maximized such that $W$ is evolved to yield the optimal performance. This subsection presents the MaOO algorithm which is considered for designing $W$ and hence, plays the central role in this transfer learning approach.

\subsubsection{Optimization Criteria}
\label{OC}
The proposed approach involves maximization of classification performance metrics. The true class labels and the predicted class labels are compared, and the corresponding counts of agreement and disagreement, generate the confusion matrix \cite{TheodoridisCh10}. Based on the confusion matrix, the following performance metrics \cite{MicroComBCI16} are obtained which attain a value of 1 for perfect classification.  
\begin{itemize}
	\item Recall
	\item Precision
	\item Accuracy
	\item F1\_score
	\item Specificity
	\item $\kappa$-coefficient
\end{itemize} 

All of the performance metrics are essential. If there is an imbalance between the classes (unequal number of instances of different classes), situations can arise where the performance metrics does not agree with each other \cite{classimbalanceprob}. For example, when number of a instances of one class (say Class 1) is much less than the number of instances of another class (say Class 2) and only the instances of Class 2 are correctly classified, then Accuracy would be very high whereas Specificity of Class 2 would be very low. Similar situations mandate to note all the performance metrics which when high imply good classification. Hence, there are six performance metrics (optimization criteria) which are to be maximized while performing classification in the projected feature space. 

It is to be noted that the performance metrics are all defined for binary classification. Hence, when there are more than two classes, these classification performance metrics should be maximized for each of the classes on a one-against-all basis which are directly attainable from the multi-class confusion matrix. Hence, as number of classes increase, number of objectives increase proportionately. However, it should be highlighted that while the number of objectives (and hence, the complexity of optimization problem) increases, yet the proposed framework can readily handle any multi-class feature extraction and classification technique.

As the criteria are in conflict with each other (i.e. maximizing one of the objectives does not necessarily mean another objective is maximized) and there are more than three objectives, this forms a problem of many-objective optimization \cite{CIEC16}.

\subsubsection{Formulation of Optimization Problem}
\label{formulation}
A many-objective optimization (MaOO) problem \cite{CIEC16,favour1} is defined as shown in Eq. \eqref{eqn: Prob1} to \eqref{eqn: Prob4} where the decision space is formed by the intersection of the constrained regions and is bounded by the lower ($x_i^L$) and upper ($x_i^U$) bounds of the decision variables ($x_i$).\\
Maximize
\begin{equation}
F(X) = [f_1(X),f_2(X),\cdots,f_M(X)]
\label{eqn: Prob1}
\end{equation}
subjected to $J$ inequality constraints (if any)
\begin{equation}
g_j(X)\geq 0,j=1,2,\cdots J
\end{equation}
and $K$ equality constraints (if any)
\begin{equation}
h_k(X)= 0,k=1,2,\cdots K
\end{equation}
where
\begin{equation}
x_i^L\leq x_i \leq x_i^U,i=1,2,\cdots,n
\label{eqn: Prob4}
\end{equation}

According to Eq. \eqref{eqn: Prob1}, the specifications of the MaOO problem for addressing the inter-subject EEG classification problem is as follows: 
\begin{itemize}
\item Based on the discussion in Section \ref{OC}, $F$ is a $6$-dimensional objective vector, i.e., $M=6$, where each of the objectives is one of the six performance metrics. 

\item Classification performance metrics are dependent on the projected feature vectors and in turn dependent on $W$. Hence, the solution vector $X$ is defined as an encoding of the elements of $W$ such that $W$ matrix can be obtained by converting the $X$ vector in row-major format. Based on the discussion in Section \ref{Theory}, the dimension of $X$ turns out as $n=D\times d$. 

\item Since, $X$ is an encoding $W$, it can be stated that $F$ is a function of $X$ as required by Eq. \eqref{eqn: Prob1}. For this application, there are no inequality ($J=0$) or equality ($K=0$) constraints. 
\end{itemize}

Dictated by its superior performance \cite{DEMO}, Differential Evolution for Multi-objective Optimization (DEMO) \cite{DEMO} is used to address the concerned BCI problem.

\subsubsection{Differential Evolution for Multi-objective Optimization (DEMO)}
This work uses Differential Evolution for Multi-objective Optimization (DEMO) to solve the concerned problem. Like the single-objective version of Differential Evolution, DEMO consists of the following steps: (a) initialization, (b) mutation, (c) recombination, and, (d) selection.

\begin{itemize}
	\item \emph{Initialization:} A population of $NP$ candidate solutions is randomly initialized. This population is represented by a matrix (order: $NP\times n$) where the $i^{\text{th}}$ row represents a solution $X_i$. For the initial generation ($G=0$), each of the $i^{\text{th}}$ candidate $X_i$ is initialized as in Eq. \eqref{eqn: init}. For this work, it was assumed that $x_j^L=0$ and $x_j^U=1$. As per convention, a variable denoted by an uppercase letter represents a vector, whereas a variable denoted by the lowercase of the same letter represents an element of the vector.
	
	\item \emph{Mutation:} For every $i^{\text{th}}$ candidate and for every generation $G$, three random indices $r_1$, $r_2$ and $r_3$ are chosen such that corresponding candidates of the population can be used to generate the donor vector $V_{i,G}$. It should be mentioned that $i$, $r_1$, $r_2$ and $r_3$ have to be mutually exclusive. Mutation is performed according to Eq. \eqref{eqn: mut}. Scale factor, $F$, used for mutation is a random number chosen between 0 and 2.
	
	\item \emph{Recombination:} For every $i^{\text{th}}$ candidate and for every generation $G$, the candidate vectors $X_{i,G}$ and the donor vectors $V_{i,G}$ are combined with a crossover rate ($CR$) to yield the trial vectors for the next generation, $U_{i,G+1}$ according to Eq. \eqref{eqn: recomb}. For a random index $I_{rand}\in{1,2,\cdots,n}$ irrespective of $CR$, $v_{iI_{rand},G}$ is selected for $u_{iI_{rand},G}$ so that $U_{i,G+1}$ is never identical to $X_{i,G}$. For this work, $CR$ is chosen as 0.8 so that the probability of selecting $v_{ij,G}$ is high.
	
	\item \emph{Selection:} For every $i^{\text{th}}$ candidate and for every generation $G$, $U_{i,G+1}$ and $X_{i,G}$ are compared and $X_{i,G+1}$ is generated according to the selection step which proceeds as Eq. \eqref{eqn: sel}. DEMO differs in this step from the single objective version of Differential Evolution, as comparison of the candidate and trial solutions cannot be done based on scalar valued fitness of the solutions. For comparison of the solution vectors (say $X_1$ and $X_2$) based on objective vectors, Pareto-dominance relation \cite{MicroComBCI16,alphaDEMO} is considered which is defined in Eq. \eqref{eqn: PD} for a maximization problem. According to Eq. \eqref{eqn: PD}, $X_1$ Pareto-dominates $X_2$, $X_1\succ X_2$, when $X_1$ is at least as good as $X_2$ in all objectives and is strictly better than $X_2$ in at least one of the objectives.
\end{itemize}
\begin{equation}
\begin{split}
x_{ij,0}=x_j^L+rand(0,1)\times(x_j^U-x_j^L)\\
\text{where }i\in\{1,2,\cdots,NP\}\text{ and }j\in\{1,2,\cdots,n\}
\label{eqn: init}
\end{split}
\end{equation}
\begin{equation}
\begin{split}
V_{i,G}=X_{r_1}+F\times(X_{r_2}-X_{r_3})\\
\forall i\in\{1,2,\cdots,NP\}
\label{eqn: mut}
\end{split}
\end{equation}
\begin{equation}
u_{ij,G+1}=
\begin{cases}
v_{ij,G} & \text{if }rand_{ij}\leq CR\text{ or }j= I_{rand}\\
x_{ij,G} & \text{if }rand_{ij}>CR\text{ and }j\neq I_{rand}
\end{cases}
\label{eqn: recomb}
\end{equation}
\begin{equation}
X_{i,G+1}=
\begin{cases}
X_{i,G} & \text{if }X_{i,G}\succ U_{i,G+1}\\
U_{i,G+1} & \text{otherwise}
\end{cases}
\label{eqn: sel}
\end{equation}
\begin{equation}
\begin{split}
\forall i\in \{1, 2, \cdots, M\}, f_i(X_1)\geq f_i(X_2)\\
\text{and } \exists j\in \{1, 2, \cdots, M\}, f_j(X_1) > f_j(X_2)
\label{eqn: PD}
\end{split}
\end{equation}

Three steps, viz., mutation, recombination, and selection continues for a predetermined number of generation ($G_{max}$) by which the algorithm is believed to have converged to the optimal solution. 

\subsubsection{Modifications of DEMO for Transfer Learning Problem}
A few modifications are done to the existing DEMO algorithm in order to make it more suitable for the current application.

\begin{enumerate}
	\item \emph{Change in selection strategy:} The problem with Pareto-dominance relation in the selection step of DEMO is described as follows. Based on the Pareto-dominance relation (Eq. \eqref{eqn: PD}), there can be three situations between two solution vectors $X_1$ and $X_2$ as follows:
	\begin{itemize}
		\item $X_1$ Pareto-dominates $X_2$, or
		\item $X_1$ is Pareto-dominated by $X_2$, or
		\item $X_1$ and $X_2$ are non-dominated by each other.
	\end{itemize}
	As $M$ increases, probability that two solutions are non-dominated with respect to each other also increases \cite{alphaDEMO,FuzzyPD}. If none of the two solution vector dominates the other, these are declared to be of the same quality which leads to poor decision-making. Hence, Pareto-dominance relation (Eq. \eqref{eqn: PD}) is a very strict condition to compare two solution vectors for MaOO problems. 
	
	Another comparatively recent approach to compare two solution vectors is by using the Favour relation \cite{favour1,favour2} which relaxes the Pareto-dominance relation as shown in Eq. \eqref{eqn: FR}. It compares two solution vectors based on the number of objectives for which one solution is better than the other. In case of a tie, the solution vectors are declared to be non-dominated. However, the probability that two solutions are non-dominated with respect to each other is much less than that in case of Pareto-dominance relation \cite{favour1,favour2}. Hence, for the concerned application, Favour relation (Eq. \eqref{eqn: FR}) is used instead of Pareto-dominance relation (Eq. \eqref{eqn: PD}) in the selection step of DEMO for comparing the trial and the candidate solutions. 
	\begin{equation}
		\begin{split}
		|\{i:f_i(X_1)>f_i(X_2),1\leq i\leq M\}|>\\
		|\{j:f_j(X_2)>f_j(X_1),1\leq j\leq M\}|
		\label{eqn: FR}
		\end{split}
	\end{equation}

	\item \emph{Ranking of candidates after selection stage:} After selection  stage, the evolved candidate solutions, within a population, are ranked based on their quality. The closeness of the objective vectors to the ideal objective vector is considered as a quality indicator. Ideal objective vector ($F_0$) is at [1, 1, 1, 1, 1, 1] which are the highest values attainable by each of the six objectives. The closer an objective vector gets to the ideal objective vector, better is the classification performance which implies better separability in the projected feature space and thus yields a better weight matrix $W$. Hence, ranking is performed based on $I_{dist}(F_i)$, which is the Euclidean distance between an objective vector ($F_i=F(X_i)$) and the ideal objective vector ($F_0$) as shown by Eq. \eqref{eqn: Idist}.
	\begin{equation}
		I_{dist}(F_i) = \sqrt{\sum_{j=1}^{M}(f_{j}(X_i)-f_{0j})^2}
		\label{eqn: Idist}
	\end{equation}
	
	It should be mentioned that $I_{dist}$ is used as a ranking strategy, but it cannot directly be used as a single objective for optimization as the optimization criteria are in conflict with each other (as discussed in Section \ref{OC}).
	
	\item \emph{Dealing with local maxima:} DEMO does not have any inherent capability to avoid local optima. Hence, if any candidate changes the value of $I_{dist}$ by less than $10^{-1}$ in 10 generations, then this implies that the candidate is approaching a maxima (local/global). If there are multiple candidates satisfying this criteria, only one candidate amongst these which has the least value of $I_{dist}$ is re-initialised (by Eq. \eqref{eqn: init}) to a random location. This provides a chance to attain a much more lower value of $I_{dist}$ and thus, helps in re-checking whether the population of candidate solutions is truly approaching the global maxima.
	
	\item \emph{Stopping condition for DEMO:} As an alternative stopping condition, to avoid unnecessary delay in the training stage of the BCI application, if the minimum value of $I_{dist}$ over the entire population changes by less than $10^{-2}$ (i.e. remains almost constant) in 50 generations, the optimization is terminated.
\end{enumerate}

Thus, DEMO, along with these modifications, results in a set of trade-off solutions from which the most preferred solution has to be chosen in order to yield the optimal $W$ for projecting the features into another dimension which will facilitate inter-subject EEG classification with satisfactory classification performance. 

\subsubsection{Decision Making}
After termination of the MaOO algorithm, the solutions in the final population which are not dominated (according to Favour relation) by any other solution of the population forms the non-dominated solution set. This is the best estimation of the Pareto-optimality \cite{alphaDEMO} as attained by the MaOO algorithm. Pareto-optimality is the trade-off state of optimization where improvement in one of the objectives is possible only at the cost of deterioration in terms of another objective. The set of objective vectors corresponding to the non-dominated solution set is called the Pareto-front ($PF$). Since a single optimal solution is not achieved through the MaOO algorithm, a strategy has to be adopted in order to choose which one of the non-dominated solutions is to be used for designing the weight matrix $W$. 

As there is no preference ordering among the objectives i.e. all the objectives are of equal priority, so the solution $X^\star$ corresponding to the objective vector $F_{opt}(X)$, which is closest to the ideal objective vector $F_0 = [1,1,1,1,1,1]$, yields the equi-best compromise among the performance metrics and hence, $X^\star$ is chosen for designing $W$. As the objective space is real-valued, for measuring the closeness of the two objective vectors Euclidean distance is used i.e. $I_{dist}$ from Eq. \eqref{eqn: Idist} is re-used. It is important to note that if one of the classes has higher importance in case of multi-class classification, then during decision making weighted distance metric should be used with higher weights for the classification performance metrics of the more preferred class.

Decision making to obtain the optimal solution ($X^\star$) proceeds as shown in Eq. \eqref{eqn: decisionmaking} after which $X^\star$ is converted into matrix (row-major format) to yield the weight matrix $W$.
\begin{equation}
\begin{split}
W \xleftarrow{\text{row-major format}} X^\star\\
\text{where, }X^\star = F_{opt}^{-1}(X)
\end{split}
\label{eqn: decisionmaking}
\end{equation}
\begin{equation*}
\text{where, }F_{opt}(X) = \argmin_{F_i\in PF}I_{dist}(F_i)
\end{equation*}

Thus, the proposed approach yields $W$ in order to project the features in to a $d$-dimensional space where source-trained target EEG classification performance is maximized. 

\subsection{Feature Transformation}
At the end of the transfer learning stage, the optimized weight matrix $W$ is used for transformation of feature matrix (Train-Set and Test-Set) according to Eq. \eqref{eqn: Proj}. The order of weight matrix is $d \times D = d \times 98$. Dimension of projected feature space ($d$) is varied in a wide range to study the sensitivity of the proposed approach towards this parameter.

\subsection{Classification}
Linear Support Vector Machine (SVM) \cite{MBEC14,conecct2014,semcco13} has two-fold purpose of acting as the fitness evaluator in the transfer learning stage and as the classifier. It is a supervised binary classifier which builds a unique hyperplane separating the two classes. The hyperplane is solely defined by the feature vectors closest to the hyperplane which are called the support vectors. This classifier tries to maximize the distance between the support vectors and the hyperplane and hence, it is also known as maximum margin hyperplane. If the classes are non-separable near the hyperplane, the margin has to be maximized yet it should be narrow such that the number of samples inside the band is minimized. A parameter regulating this trade-off is a positive constant $C$ (smoothing parameter). After testing with $C = [0.01, 0.1, 0.5, 1, 5, 10, 20]$, best performance is obtained for $C=1$.

\section{Results}
 \label{RnD}
This section presents the performance of the proposed approach on BCI Competition III - Data set IVa. The work is executed on a computer having 4GB RAM and Intel Core i3 @ 2.3GHz processor running the 32-bit version of MATLAB R2012b. The experiment is repeated 50 times, with the same proportion but different randomization at cross-validation phase, for each pair of source and target data sets and the results are noted.

\subsection{Parameters of Optimization Algorithm}
For implementing DEMO in the proposed framework, the sensitivity of the parameters has been studied and the best training performance is observed for values which are mentioned in Table \ref{tab:Parameters} i.e. Test-Set is not used in determining these parameters.

\begin{center}
	\begin{table}[h]
		\caption{Values of Different Parameters used in this work.}
		\begin{center}
		\resizebox{0.6\columnwidth}{!}{
			\begin{tabular}{l|l|l}
				\label{tab:Parameters}
				\textbf{Parameters}    & \textbf{Explanation} & \textbf{Values}\\
				\hline\hline
				$NP$          & Population size            & 100 \\
				\hline
				$G_{max}$     & Maximum generations        & 2000\\
				\hline
				$F$           & Scale Factor               & Randomly chosen \\
				&                            & between 0 and 2 \\
				\hline
				$CR$          & Crossover Rate             & 0.8\\
				\hline
				$d$			  & Dimension of projected space & 2\\
				\hline
			\end{tabular}
		}
	\end{center}
	\end{table}
\end{center}

Changing of the parameter values from the recommended settings has the following effects:
\begin{enumerate}
	\item Increasing $NP$ with constant $G_{max}$, does not improve the performance. This is because proportionately more number of times random initialization of candidates are occurring rather than mutation and recombination operations. This, in turn, implies lesser number of good solutions are being transmitted. 
	
	\item As will be evident from subsequent sections, the optimization algorithm terminates much earlier than reaching $G_{max}$. Thus, keeping $NP$ fixed while increasing $G_{max}$, does not cause any further improvement in the results.
	
	\item The parameter $CR$ controls how far from the parent candidate, a solution is generated. Hence, it is usually kept high as a new candidate far from the parent solution would give better exploration of the search space. However, when $CR > 0.8$, the trial vector becomes mostly independent of the parent candidate, which leads to poor performance.
	
	\item The sensitivity of the proposed approach towards the dimension of the project space ($d$) is presented separately in the subsequent sections.   
\end{enumerate} 

Using the parameters mentioned in Table \ref{tab:Parameters}, the classification results are noted in this manuscript.

\begin{center}
	\begin{table*}[!htbp]
		\caption{Comparing the Performance (\%) of Proposed Transfer Learning Approach with Other Relevant Approaches.}
		\resizebox{\textwidth}{!}{
			\begin{tabular}{c|c|c|c|c|c|c|c|c|c|c|c|c|c}
				\label{tab:Results}
				\textbf{Source} & \textbf{Target} & \multicolumn{3}{c|}{\textbf{Accuracy}} & \multicolumn{3}{|c|}{\textbf{F1\_score}} & \multicolumn{3}{|c|}{\textbf{Specificity}} & \multicolumn{3}{|c}{\textbf{Testing Time (ms)}}\\
				\cline{3-14}
				\textbf{Subject ID} & \textbf{Subject ID} & \textbf{Proposed} & \textbf{TSM} & \textbf{EER} & \textbf{Proposed} & \textbf{TSM} & \textbf{EER} & \textbf{Proposed} & \textbf{TSM} & \textbf{EER} & \textbf{Proposed} & \textbf{TSM} & \textbf{EER}\\
				& & \textbf{Approach} & & & \textbf{Approach} & & & \textbf{Approach} & & & \textbf{Approach} & & \\
				\hline\hline
				aa & aa & 66.19 & \textbf{68.57} & 65.36 & 63.96 & \textbf{68.12} & 67.34 & \textbf{72.38} & 70.00 & 59.29 & 17.3 & \textbf{10.3} & 13.5 \\
				\hline
				aa & al & \textbf{91.90} & 89.64 & 51.79 & \textbf{91.54} & 89.61 & 65.82 & \textbf{96.19} & 90.00 & 10.71 & 16.3 & \textbf{9.00} & 9.40 \\
				\hline
				aa & av & \textbf{60.95} & 59.29 & 50.36 & 60.95 & 60.96 & \textbf{66.67} & \textbf{60.95} & 55.00 & 1.43 & 14.8 & \textbf{9.10} & 9.20 \\
				\hline
				aa & aw & \textbf{75.71} & 63.21 & 49.64 & \textbf{75.83} & 63.08 & 65.53 & \textbf{75.24} & 63.57 & 3.57 & 16.4 & 9.60 & \textbf{9.20} \\ 
				\hline
				aa & ay & \textbf{70.48} & 68.21 & 58.57 & 65.56 & 67.64 & \textbf{68.31} & \textbf{84.76} & 70.00 & 27.86 & 16.1 & \textbf{9.20} & \textbf{9.20} \\
				\hline

				al & aa & \textbf{59.05} & 57.50 & 46.43 & 57.00 & 57.35 & \textbf{57.63} & \textbf{63.81} & 57.86 & 20.00 & 23.6 & 9.80 & \textbf{9.40} \\
				\hline
				al & al & 94.29 & \textbf{95.71} & 93.21 & 94.12 & \textbf{95.80} & 93.60 & \textbf{97.14} & 93.57 & 87.14 & 17.9 & \textbf{9.30} & 9.60 \\
				\hline
				al & av & \textbf{62.86} & 61.79 & 60.36 & 60.20 & \textbf{63.73} & 61.05 & \textbf{69.52} & 56.43 & 58.57 & 16.3 & \textbf{9.40} & 9.50 \\
				\hline
				al & aw & \textbf{84.76} & 75.00 & 65.71 & \textbf{84.31} & 75.18 & 49.47 & 87.62 & 74.29 & \textbf{97.86} & 14.2 & 8.90 & \textbf{8.80} \\
				\hline
				al & ay & \textbf{85.24} & 80.00 & 51.43 & \textbf{84.58} & 79.71 & 10.53 & 89.52 & 81.43 & \textbf{97.14} & 13.9 & \textbf{9.30} & 9.50 \\
				\hline

				av & aa & 57.62 & \textbf{58.57} & 53.57 & 42.58 & \textbf{57.35} & 19.75 & 83.81 & 61.43 & \textbf{95.71} & 14.8 & \textbf{9.60} & 9.80 \\
				\hline
				av & al & 93.33 & \textbf{95.00} & 83.93 & 82.23 & \textbf{95.17} & 85.25 & 89.52 & \textbf{91.43} & 75.00 & 14.7 & \textbf{9.10} & 10.00 \\
				\hline
				av & av & 65.71 & \textbf{68.57} & 62.50 & 65.38 & \textbf{70.07} & 66.88 & \textbf{66.67} & 63.57 & 49.29 & 18.5 & \textbf{9.40} & \textbf{9.40} \\
				\hline
				av & aw & \textbf{80.00} & 76.43 & 63.93 & \textbf{76.40} & 76.26 & 60.39 & \textbf{95.24} & 77.14 & 72.86 & 14.4 & 9.70 & \textbf{9.30} \\
				\hline
				av & ay & \textbf{74.29} & 62.50 & 54.29 & \textbf{70.00} & 62.09 & 68.63 & \textbf{88.57} & 63.57 & 8.57 & 17.7 & \textbf{9.20} & \textbf{9.20} \\
				\hline

				aw & aa & \textbf{60.95} & 56.79 & 47.86 & \textbf{60.95} & 57.54 & 59.89 & \textbf{60.95} & 55.00 & 17.86 & 20.8 & \textbf{9.40} & 9.80 \\
				\hline
				aw & al & 90.48 & \textbf{90.71} & 82.14 & 90.57 & \textbf{90.85} & 84.85 & \textbf{89.52} & 89.29 & 64.29 & 20.8 & \textbf{9.00} & 9.30 \\
				\hline
				aw & av & \textbf{61.90} & 57.86 & 53.57 & 66.94 & 60.14 & \textbf{68.29} & 46.67 & \textbf{52.14} & 7.14 & 20.8 & \textbf{9.40} & 10.00 \\
				\hline
				aw & aw & 88.10 & \textbf{92.86} & 83.93 & 88.48 & \textbf{92.81} & 84.32 & 84.31 & \textbf{93.57} & 81.43 & 20.8 & 9.30 & \textbf{9.10} \\
				\hline
				aw & ay & \textbf{84.76} & 57.14 & 68.57 & \textbf{84.31} & 58.62 & 75.00 & \textbf{87.62} & 53.57 & 42.86 & 20.8 & \textbf{9.00} & 9.70 \\
				\hline

				ay & aa & \textbf{63.81} & 55.36 & 48.93 & 55.81 & \textbf{56.45} & 49.47 & \textbf{81.90} & 52.86 & 47.86 & 15.2 & \textbf{9.60} & 9.80 \\
				\hline
				ay & al & \textbf{97.62} & 93.57 & 92.14 & 86.73 & \textbf{93.71} & 92.14 & \textbf{94.29} & 91.43 & 92.14 & 13.9 & \textbf{9.10} & 9.30 \\
				\hline
				ay & av & 60.00 & 58.21 & \textbf{62.14} & 65.85 & 60.34 & \textbf{69.71} & 42.86 & \textbf{52.86} & 37.14 & 16.8 & \textbf{9.10} & \textbf{9.10} \\
				\hline
				ay & aw & \textbf{81.90} & 66.79 & 55.00 & \textbf{81.73} & 66.67 & 18.18 & 82.86 & 67.14 & \textbf{100.00} & 14.8 & 9.40 & \textbf{8.90} \\
				\hline
				ay & ay & 84.29 & \textbf{90.36} & 87.50 & 82.54 & \textbf{90.18} & 87.89 & \textbf{94.29} & 92.14 & 84.29 & 17.5 & \textbf{9.30} & \textbf{9.30} \\
				\hline

				\multicolumn{2}{c|}{\textbf{Mean}} & 75.85 & 71.99 & 63.71 & 73.54 & 72.38 & 63.48 & 79.47 & 70.77 & 72.29 & 16.6 & 9.30 & 9.57 \\
				\hline
				\multicolumn{2}{c|}{\textbf{Standard Deviation}} & 13.16 & 14.56 & 14.83 & 13.54 & 14.23 & 30.43 & 15.23 & 15.33 & 28.01 & 2.60 & 0.31 & 0.88 \\
				\hline
			\end{tabular}
		}
	\end{table*}
\end{center}

\subsection{Performance of the Proposed Approach}
The classification results in terms of Accuracy, F1-score, Specificity and testing time (per EEG instance) obtained using the proposed approach for all 25 combinations of source and target subjects, are mentioned in Table \ref{tab:Results}. The testing time is the total time taken by a single sample of the Test-Set (target data) for preprocessing, feature extraction, projecting the feature matrix in to the transformed space and testing the trained classifier using transformed features. Due to space constraint, the values of Recall, Precision and $\kappa$-coefficient could not be shown. These results along with the performance of the MaOO algorithm are available at \url{http://worksupplements.droppages.com/tlbci.html}. 

The values mentioned in Table \ref{tab:Results} are the average results obtained after 50 executions of the proposed approach. In order to establish the efficacy of the proposed approach, its performance is compared with Riemannian Geometry \cite{RM} and Extreme Energy Ratio \cite{BSPC15} based approaches in Table \ref{tab:Results}. These methods are used as recommended in the referenced papers and these are chosen for comparison due to the following reasons:
\begin{enumerate}
	\item Recent studies have shown Riemannian geometry \cite{RM} based techniques to perform well for inter-subject transfer. Tangent Space Mapping (TSM) \cite{RM}, which is a Riemannian geometry based method, is used on preprocessed EEG data for feature extraction and results are noted while following conventional supervised learning paradigm with TSM based features from the same dataset.
	\item The most commonly used feature in the transfer learning literature of BCI is CSP \cite{jour36,proc15,jour35}. Extreme Energy Ratio (EER) has been shown to be equivalent to CSP \cite{EER}. Hence, results from supervised learning paradigm with EER based features on the same dataset has also been noted for comparison against the proposed approach.
\end{enumerate}

It should be mentioned that during comparison with the existing methods, which are not using the proposed transfer learning approach, i.e. for supervised learning using TSM based and EER based features in Table \ref{tab:Results}, the same Train-Set and Test-Set, as defined in Table \ref{tab:HoldOut}, are used for the purposes of training and testing, respectively. 

\subsection{Comparison against inter-subject EEG classification without transfer learning}
\begin{figure}[ht]
	\centering
	\includegraphics[width = \linewidth,bb = 70 0 790 380, clip = true]{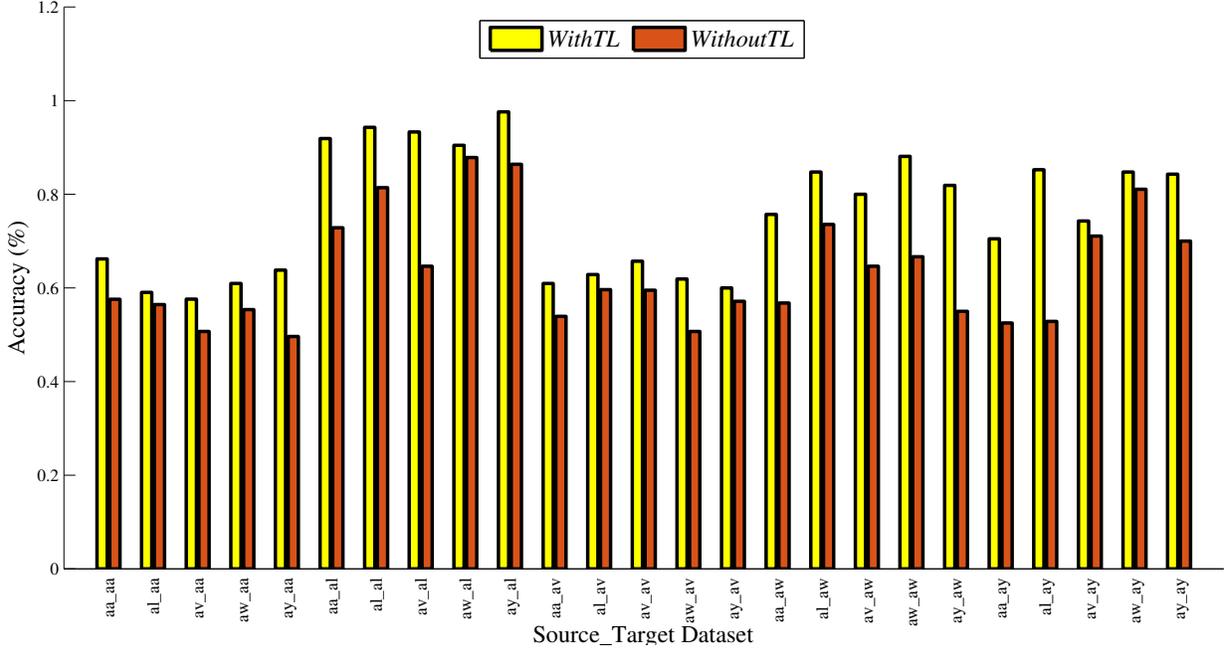}
	\caption{Comparison of Accuracy with and without Transfer Learning.}	
	\label{fig:bargraph}
\end{figure}

In order to understand the improvement in classification achieved by using the transfer learning (\emph{WithTL}) approach, the results are compared with the approach that does not use transfer learning i.e. supervised learning using PSD-estimates as features and SVM as classifier (\emph{WithoutTL}) for the same dataset. However, due to space constraint, only accuracy is compared in Fig. \ref{fig:bargraph}. The source and the target subject are mentioned in the format $src\_tar$ where $src$ is one of the five source subject\textquoteright s ID and $tar$ is one of the five target subject\textquoteright s ID as mentioned in Section \ref{dataset}. During these experiments, the specific cross-validation ratios or the portion of dataset used for training and testing are as follows:

\begin{enumerate}
	\item When source and target subjects are different i.e. $src$ is different from $tar$, the results from the \emph{WithoutTL} technique in Fig. \ref{fig:bargraph} are noted using the same Train-Set and Test-Set, as defined in Table \ref{tab:HoldOut}, for the purposes of training and testing, respectively.
	\item When source and target subjects are identical i.e. $src$ is same as $tar$, the results from the \emph{WithTL} technique in Fig. \ref{fig:bargraph} are noted using a 25-75 hold-out cross-validation such that Validation-Source-Set and Train-Set are identical to Validation-Target-Set with randomly chosen 70 samples from target EEG dataset and the remaining 210 samples for the Test-Set. This is done because of the assumption that only small amount of data from the target subject is available for training.
	\item During comparison, in order to maintain the same basis of availability of less data from the target subject, when source and target subjects are identical i.e. $src$ is same as $tar$, the results for the \emph{WithoutTL} technique in Fig. \ref{fig:bargraph} are noted using the Validation-Target-Set as the Train-Set and the Test-Set (from Table \ref{tab:HoldOut}) for training and testing, respectively.
\end{enumerate}

\subsection{Sensitivity with Dimension of the Projected Space}
The main task of the proposed approach is to determine some weights which linearly projects the feature space in to a $d$-dimensional space in order to facilitate inter-subject EEG classification. 

\begin{figure}[ht]
	\centering
	\includegraphics[width = 0.65\linewidth, bb = 20 30 450 450, clip = true]{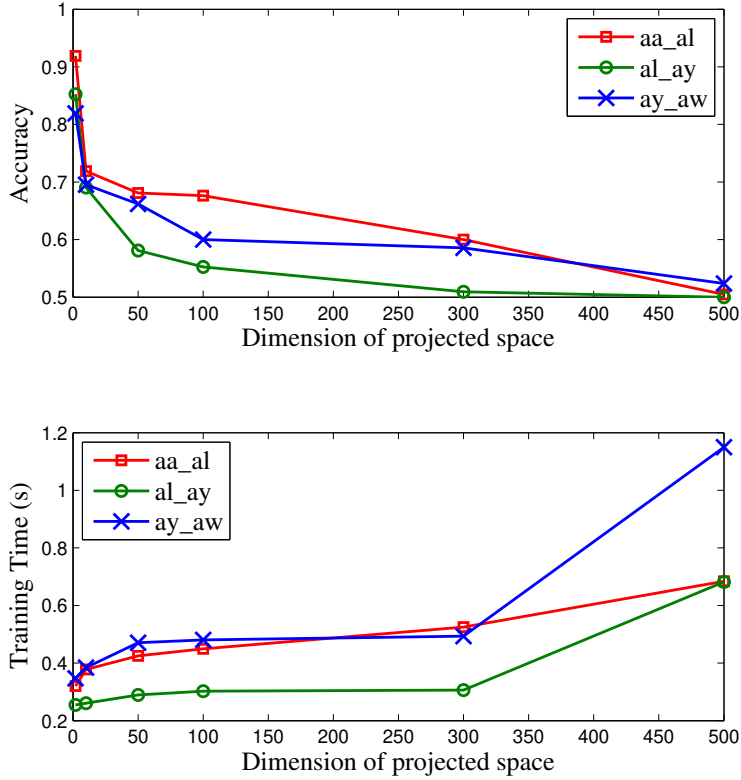}
	\caption{Variation of Accuracy and Training Time with Dimension of Projected Space ($d$).}
	\label{fig:vary_d}
\end{figure}

The variation of accuracy and training time with $d$ is noted in Fig. \ref{fig:vary_d} at $d = \{2, 10, 50, 100, 300, 500\}$ with respect to three source-target combinations. These three combinations are the three cases for which best learning (measured from Fig. \ref{fig:bargraph} by taking difference between accuracies with and without transfer learning) has been observed. Accuracy decreases as $d$ increases with highest accuracy at $d=2$ and lowest accuracy of 50\% at $d=500$ representing random classification. Training time increases with increase in $d$ as learning more number of weights require more time for each step of DEMO. Thus, $d=2$ has been proposed in Table \ref{tab:Parameters}.

\subsection{Maximization of Optimization Criteria}
\begin{figure}[ht]
	\centering
	\includegraphics[width = 0.8\linewidth, bb = 70 40 630 470, clip = true]{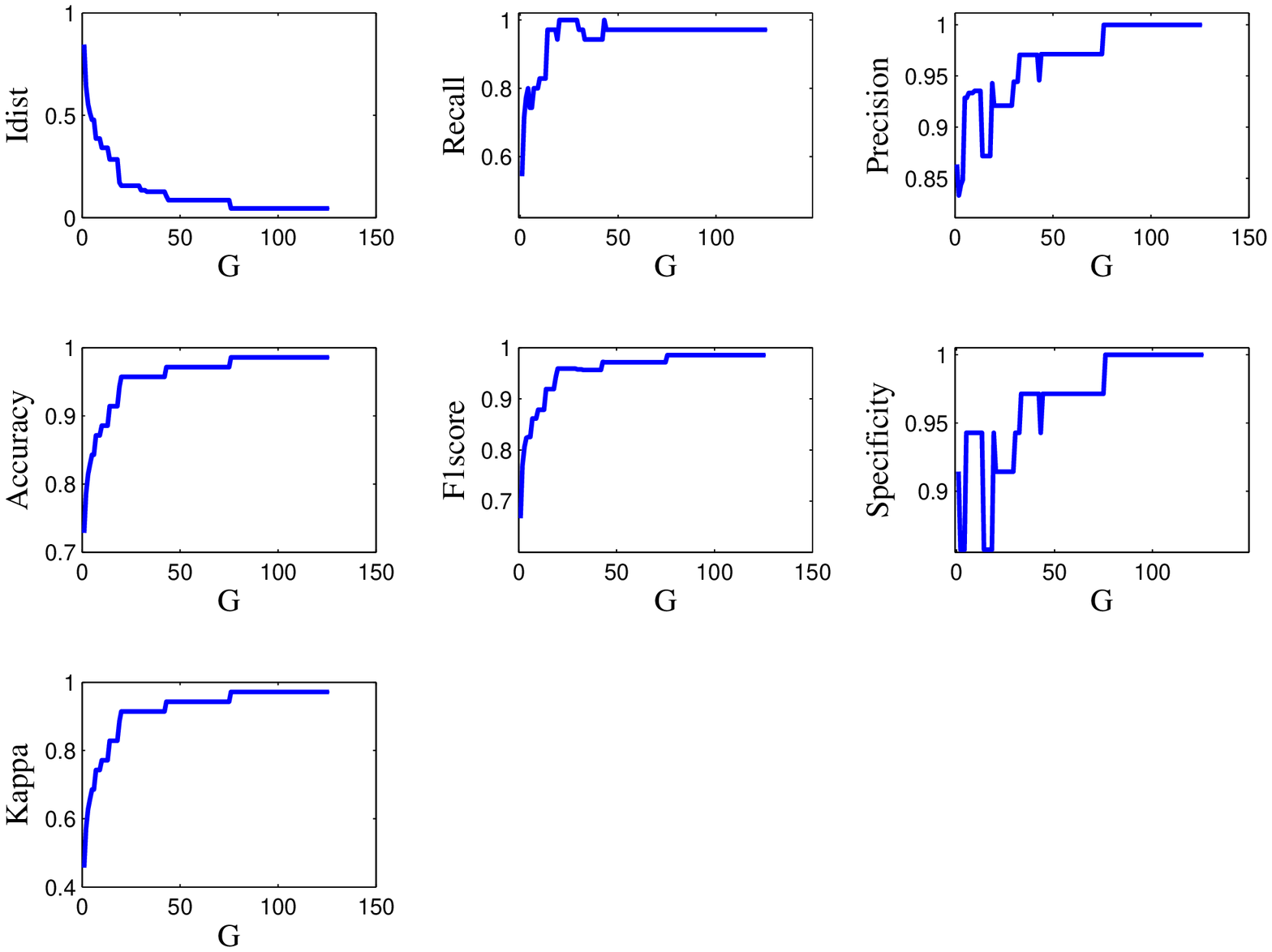}
	\caption{Improvement of Fitness of MaOO Model of Transfer Learning.}
	\label{fig:MaOO performance}
\end{figure}

Maximization of the six objectives (classification performance metrics) with respect to generations ($G$) is shown in Fig. \ref{fig:MaOO performance}. Minimization of the distance between nearest point of approximated $PF$ and $F_0$ (denoted as $Idist$) with respect to $G$ is also shown in Fig. \ref{fig:MaOO performance}. These results are shown for $d=2$ where accuracy is maximum and for a single run of the proposed approach for the source-target subject pair: $aa\_al$. 

\subsection{Distribution of the Classes in Projected Space}
\begin{figure}[ht]
	\centering
	\includegraphics[width = 0.70\linewidth, bb = 50 20 590 460, clip = true]{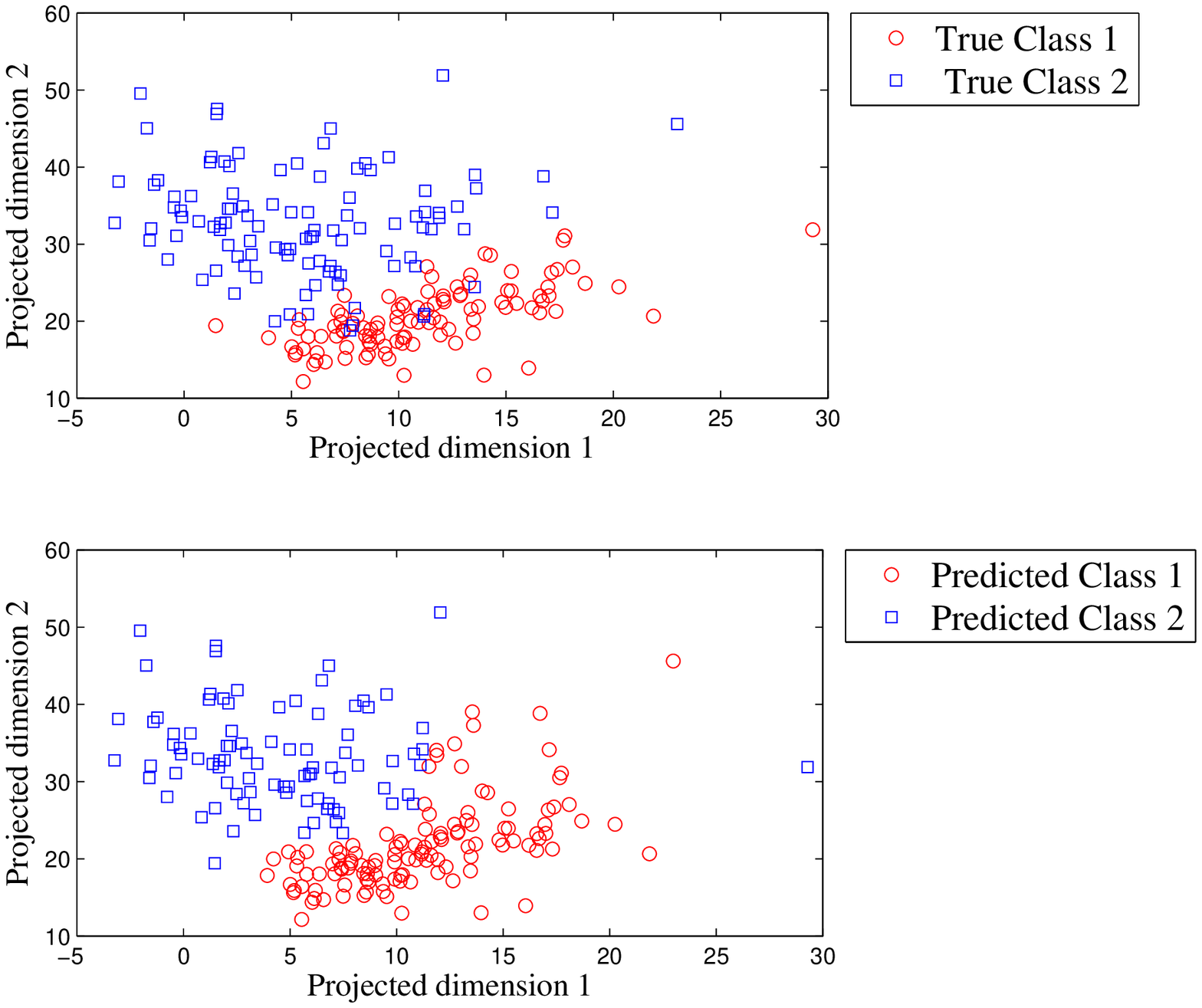}
	\caption{Comparison of True Samples with the Predicted Samples using the Proposed Approach.}
	\label{fig:classes}
\end{figure}

Classified samples are compared with the true samples in the projected feature space (for $d=2$) in Fig. \ref{fig:classes} which shows most of the samples are successfully identified. Observations made in Fig. \ref{fig:classes} are for a single run of the proposed approach and for the source-target subject pair: $aa\_al$. 

\subsection{Statistical Analysis}
For statistical validation of the results, paired $t$-test, Wilcoxon Signed Rank test \cite{WSRT} and Holm-Bonferroni test \cite{HBT} are performed. The tests are performed with $25$ source-target pairs, under the following assumptions of null ($H_0$) and alternate ($H_a$) hypotheses:

$H_0$: No significant change in performance is noted.

$H_a$: Better performance is not a chance event.

\subsubsection{Paired $t$-Test}
For statistical validation of these results (accuracy), paired $t$-tests are performed to compare the proposed transfer learning approach against (a) PSD based supervised learning (the framework without using transfer learning i.e. \emph{WithouTL}), (b) TSM based supervised learning (a recent transfer learning approach) and (c) EER based supervised learning (most commonly used features for BCI problems). Resultant $t$-statistics are noted in Table \ref{tab:StatAna}. All the values indicate that the tests reject $H_0$ at $95\%$ confidence interval and thereby, validating the claim of better performance of the proposed approach ($H_a$).

\subsubsection{Wilcoxon Signed Rank Test}
This is a nonparametric test, similar to the paired $t$-test, which test whether two dependent group of samples were drawn from the populations having same distributions \cite{WSRT}. In order to perform this test, the pairs of results (accuracy) from two approaches are ranked in ascending order of absolute difference and these ranks are labelled using the sign of the difference (hence, called signed ranks). For ties during ranking, each of the position is assigned a rank equal to the average of the ranks spanned by them. The $W$-statistics is given by the minimum of the sum of positive ranks and negative ranks. If number of pairs ($n_{obs}$) are large like 20 or more, the distribution can be approximated as normal with mean ($\mu_W$) and standard deviation ($\sigma_W$) as shown by Eq. \eqref{eqn:Wstat}. The signed ranks and $W$-values are noted in Table \ref{tab:StatAna}. For all the comparisons, $H_0$ are rejected as $W$-value is less than the critical value of $W_{0.05,25}=89$.
	\begin{equation}
	\begin{split}
	&\mu_W =\frac{n_{obs}(n_{obs}+1)}{4}\text{  and  }\\
	&\sigma_W=\sqrt{\frac{n_{obs}(n_{obs}+1)(2n_{obs}+1)}{24}}
	\end{split}
	\label{eqn:Wstat}
	\end{equation}

\subsubsection{Holm-Bonferroni Test}
To control the errors arising out of multiple comparisons, this post-hoc test \cite{HBT} is performed. Applying Bonferroni correction to the significance level of each hypotheses deals with the family-wise error rate. The $p$-values from the proposed approach and its competitors are listed under paired $t$-test and Wilcoxon Signed Rank test in Table \ref{tab:StatAna}. Within each family consisting of pairs of comparison, the hypotheses are ranked with lowest $p$-value having a numeric rank of 1 and so on as shown in Table \ref{tab:StatAna} as $R_{H,i}$.  Using this rank, the significance level for 95\% confidence interval is adjusted to $\alpha'$ using Eq. \eqref{eqn:HBT} where $k_a$ is number of constituents of the family. The adjusted value of $\alpha'$ is noted in Table \ref{tab:StatAna}. All those hypotheses are rejected for which $p$-values are smaller than the corresponding $\alpha'$, and the test fails to reject the rest of the hypotheses.
	\begin{equation}
	\alpha'=\frac{0.05}{(k_a+1-R_{H,i})}
	\label{eqn:HBT}
	\end{equation}

\begin{table*}[!htbp]
	\begin{center}
		\caption{Validating Proposed Approach via Paired $t$-Test (PTT), Wilcoxon Signed Rank Test (WSRT) and Holm-Bonferonni Test (HBT).}
		\resizebox{\textwidth}{!}{
			\begin{threeparttable}
				\begin{tabular}{c|c|c|c|c|c|c|c}
					\textbf{Source} & \textbf{Target} & \multicolumn{3}{c|}{\textbf{Comparing Accuracy for PTT against}} & \multicolumn{3}{|c}{\textbf{Signed Rank for WSRT against}}\\
					\cline{3-8}
					\textbf{Subject ID} & \textbf{Subject ID} & \textbf{PSD approach} & \textbf{TSM approach} & \textbf{EER approach} & \textbf{PSD approach} & \textbf{TSM approach} & \textbf{EER approach} \\
					\hline\hline
					aa & aa & 0.0862 & -0.0238 & 0.0083 & 11 & -11 & 1 \\
					\hline
					al & aa & 0.0262 & 0.0155 & 0.1262 & 1.5 & 5 & 15 \\
					\hline
					av & aa & 0.0691 & -0.0095 & 0.0405 & 9 & -2 & 7 \\
					\hline
					aw & aa & 0.0559 & 0.0416 & 0.1309 & 7 & 16 & 16 \\
					\hline
					ay & aa & 0.1417 & 0.0845 & 0.1488 & 16 & 20 & 17 \\
					\hline
					
					aa & al & 0.1904 & 0.0226 & 0.4011 & 21 & 9 & 25 \\
					\hline
					al & al & 0.1286 & -0.0142 & 0.0108 & 15 & -4 & 2 \\
					\hline
					av & al & 0.2869 & -0.0167 & 0.0940 & 24 & -7 & 12 \\
					\hline
					aw & al & 0.0262 & -0.0023 & 0.0834 & 1.5 & -1 & 11 \\
					\hline
					ay & al & 0.1119 & 0.0405 & 0.0548 & 13 & 15 & 9 \\
					\hline
					
					aa & av & 0.0702 & 0.0166 & 0.1059 & 10 & 6 & 13 \\
					\hline
					al & av & 0.0322 & 0.0107 & 0.0250 & 4.5 & 3 & 4 \\
					\hline
					av & av & 0.0619 & -0.0286 & 0.0321 & 8 & -12 & 5.5 \\
					\hline
					aw & av & 0.1119 & 0.0404 & 0.0833 & 13 & 14 & 10 \\
					\hline
					ay & av & 0.0286 & 0.0179 & -0.0214 & 3 & 8 & -3 \\
					\hline
					
					aa & aw & 0.1892 & 0.1250 & 0.2607 & 20 & 23 & 22 \\
					\hline
					al & aw & 0.1119 & 0.0976 & 0.1905 & 13 & 21 & 20 \\
					\hline
					av & aw & 0.1536 & 0.0357 & 0.1607 & 18 & 13 & 18 \\
					\hline
					aw & aw & 0.2143 & -0.0476 & 0.0417 & 22 & -17 & 8 \\
					\hline
					ay & aw & 0.2690 & 0.1511 & 0.2690 & 23 & 24 & 23 \\
					\hline
					
					aa & ay & 0.1798 & 0.0227 & 0.1191 & 19 & 10 & 14 \\
					\hline
					al & ay & 0.3238 & 0.0524 & 0.3381 & 25 & 18 & 24 \\
					\hline
					av & ay & 0.0322 & 0.1179 & 0.2000 & 4.5 & 22 & 21 \\
					\hline
					aw & ay & 0.0369 & 0.2762 & 0.1619 & 6 & 25 & 19 \\
					\hline
					ay & ay & 0.1429 & -0.0607 & -0.0321 & 17 & -19 & -5.5 \\
					\hline
					\multicolumn{2}{c|}{\textbf{Paired $t$-Test (PTT) \tnote{a}}} & $7.1582$, $<10^{-5}$, R & $2.6552$, $0.0138$, R & $5.5298$, $1.1\times 10^{-5}$, R & - & - & - \\
					\hline
					\multicolumn{2}{c|}{\textbf{Wilcoxon Signed Rank Test (WSRT) \tnote{b}}} & - & - & - & $325$, $0$, $0$, $0$, R & $252$, $73$, $73$, $0.0160$, R & $316.5$, $8.5$, $8.5$, $0$, R \\
					\hline
					\multicolumn{2}{c|}{\textbf{Holm-Bonferroni Test (HBT) \tnote{c}}} & $1$, $0.0167$, R & $3$, $0.0500$, R & $2$, $0.0250$, R & $1$, $0.0167$, R & $2$, $0.0250$, R &  $1$, $0.0167$, R\\
					\hline
				\end{tabular}
				\begin{tablenotes}
					\item[a]{For PTT: value from $t$-statistics, $p$-value, Acceptance(A) or Rejection(R) of $H_{0,i}$ at 95\% confidence interval}
					\item[b]{For WSRT: Sum of positive ranks, Sum of negative ranks, value from $W$-statistics, $p$-value, Acceptance(A) or Rejection(R) of $H_{0,i}$ at 95\% confidence interval}
					\item[c]{For HBT: Rank ($R_{H,i}$), Bonferroni corrected significance level ($\alpha'$) for 95\% confidence interval, Acceptance(A) or Rejection(R) of $H_{0,i}$}
				\end{tablenotes}
			\end{threeparttable}
			\label{tab:StatAna}
		}
	\end{center}
\end{table*}

\section{Discussions}
 \label{Discuss}
In the previous sections, the proposed framework for transfer learning has been applied to BCI Competition III Data set IVa. Its performance has been noted and compared other standard classification techniques and transfer learning methods. A discussion on the performance analysis of the proposed work is presented in this Section. The major observations from the results reported in Section \ref{RnD} are as follows: 
 
\begin{enumerate}
	\item \emph{Comparison of the classification performance:} Based on the performance observed in Table \ref{tab:Results}, it can be argued that the proposed approach improves over the approaches with which the method is compared for most of the $25$ cases. For the cases where the other approaches have outperformed the proposed approach, it should be noted that the difference in performance values is very small from those obtained by the next best approach. Hence, the performance of the proposed work is at least as good as these competitive approaches. The reason for its better performance can be credited to the fact that the framework directly trains the classifier in order to maximize the classification performance.
	
	\item \emph{Comparison of the testing time:} In terms of testing time, the proposed work takes much more time than the other approaches. It should be noted that a feature transformation phase is required in addition to other phases for the proposed work. Hence, it is essential to use a smaller value of $d$ because when $d$ is small, lesser time is spent on projecting the features (as lesser number of multiplications are required) and hence, increases the real-time applicability of the work in BCI domain. 
	
	\item \emph{Efficacy of the proposed work:} A very important observation from Fig. \ref{fig:bargraph} is that for any given target subject, the accuracy achieved by \emph{WithoutTL} method is outperformed by the accuracies achieved by \emph{WithTL} method using any source subject. As these accuracy values are not a chance event (statistically validated), hence, it can be stated that the primary goal of the proposed work of presenting a means to efficiently perform single-source trained target EEG classification, has been achieved successfully.
	
	\item \emph{Swiftly approaching convergence:} The take-away from the plots in Fig. \ref{fig:MaOO performance} is that DEMO convergences by $120$ generations. Hence, choosing $G_{max}=2000$ is sufficient for the optimization problem as mentioned in Table \ref{tab:Parameters}.
	
	\item \emph{Handling inter-class overlap around decision boundary:} As noted from Fig. \ref{fig:classes}, misclassification is observed near the decision boundary. This is because a linear classifier (SVM) is used and true samples are seen to overlap (i.e. inter-class overlapping exists) around decision boundary. By tuning the smoothing parameter such misclassification has been reduced.
	
	\item \emph{Reasons for using multiple performance measures:} As mentioned earlier, in order to address class imbalance problem, all the classification performance measures are needed. Another classification measure, which single handedly takes care of class imbalance, is area under the receiver operating characteristics (or AUC). However, in order to plot the receiver operating characteristics (ROC) and to evaluate the AUC from the ROC, the classifier has to be re-trained using various decision boundaries. This will lead to a computationally costly optimization criteria. Hence, the classification performance measures are used instead, which can be easily obtained from the confusion matrix.
	
	\item \emph{Reasons for not using combination of multiple supervised learning techniques:} Combining relevant information from multiple means of feature extraction techniques and predicting motor imagery tasks based on majority decision taken by an ensemble of varied classifiers \cite{R3_6_39} are other methods which can be integrated in the proposed transfer learning framework. However, as feature extraction and classification phases are involved during the testing stage, the computational time will increase for combining the information (multiple features with increased dimension and multiple predictions) and thereby, undermine the real-time applicability of the proposed work in BCI domain. Moreover, for stand-alone systems, where processing resources are scarce, such methods are computationally expensive to implement. 
	
	A recent approach using ensemble methods \cite{newlit6} builds classifiers from individual source subjects. Our proposed work tries to avoid this requirement of data from multiple source subjects. Another ensemble framework has been earlier proposed in literature \cite{R3_6_39}, it mentions the performance only in terms of classification accuracy. The testing time requirement has not been specified which is very much essential to assess the applicability of the work for real world problems. 
	
	\item \emph{Comparison of classification accuracy with the state-of-the-art approaches:} As per the author's awareness, a method involving inter-subject EEG classification using a single source subject is not available in the literature. Hence, the proposed approach cannot be compared with the existing works at an algorithmic level. However, as most of the previous works \cite{newlit4,jour36,jour35,jour34}, on transfer learning in BCI, have experimented their proposed method on the same BCI Competition III Dataset IVa as done in this paper, the accuracy of the proposed method is compared (in Fig. \ref{fig:Compare}) with the highest average classification accuracies across all the five subjects as reported in the respective papers. Also, the accuracy from the TSM based transfer learning method is noted for comparison. From the bar graph, we can conclude that the proposed with has outperformed most of the works.
	
	\begin{figure}[ht]
		\centering
		\includegraphics[width = 0.6\linewidth, bb = 25 0 510 220, clip = true]{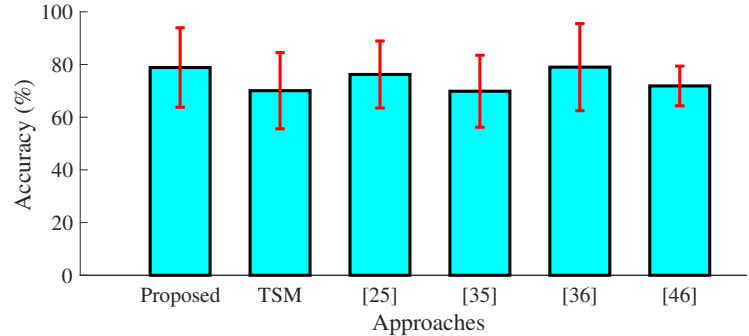}
		\caption{Comparison of the proposed transfer learning framework with state-of-the-art transfer learning approaches.}	
		\label{fig:Compare}
	\end{figure}
	
	Some other recent works in towards the concerned BCI problem and their limitations are as follows:
	\begin{itemize}
		\item A new domain adaptation technique called adaptive subspace feature matching \cite{newlit1} is used for emotion recognition and has achieved a classification accuracy of 80.46\% with 6.84\% standard deviation for offline classification. This work has been experimented only for session-to-session transfer across different sensors. Moreover, being domain adaptation technique, this method suffers from the drawback of high level of additional computation which makes in unsuitable for real-time application and stand-alone systems. These limitations are outweighed by the merits of our proposed framework.
		
		\item Another work \cite{newlit2} in emotion recognition uses conditional transfer learning for positive transfer i.e. a subject selection step precedes the transfer learning step for assessing the degree of subject transferability. However, the work only reports an accuracy of around 65\% which is outperformed by our proposed method. 
		
		\item Instead of transfer learning, another approach to handle the problem of supervised learning in presence of small amount of training data is developing methods to reduce the calibration time. A recent effort \cite{newlit5} in this direction using CSP and LDA is to address the problem by artificially generating samples to increase the size of training dataset. However, for motor imagery EEG classification, with about 25\% of training data from the subject, this method achieves an accuracy less than 75\%. Although using our proposed framework requires assistance (training data) from another subject (source subject), but based on accuracy, our proposed method is still better in similar situations.
	\end{itemize}
		
\end{enumerate}

Besides the results discussed in this section validating the superior performance of the proposed work, its merits as discussed in Section \ref{merits} enhance its value for the researchers of BCI community.

\section{Conclusion}
 \label{Con}
Due to high variance among EEG signals obtained from different subjects for the same task, a challenge posed by the BCI systems is to classify someone\textquoteright s EEG signal using a classifier trained by someone else\textquoteright s EEG signal. This situation is very serious when a differently-abled person, called the target subject, cannot provide sufficient training data and BCI system obtains training data from another unrelated person, called the source subject. This work proposes a framework to tackle this issue of inter-subject EEG classification. Raw EEG data is acquired from BCI Competition III Data set IVa from which power spectral density based features are extracted. A weight matrix is designed using information from source EEG and small amount of information from target EEG by maximizing classification performance metrics. This weight matrix is used for linear projection of feature matrix into another dimension where the non-stationarities are reduced and thereby, making classification feasible in the projected feature space with sufficient discrimination.
 
Performance analysis indicate that the proposed work can successfully address the inter-subject EEG classification problem. However, the optimization model of the transfer learning approach uses classification performance metrics and hence, is dependent on the class labels of the EEG instances (supervised). In future, the authors will be working on information theoretic metrics for the optimization criteria such that the transfer learning approach can be performed using unlabelled source and target EEG instances (unsupervised). Moreover, due to the limitation of transductive transfer learning, training phase has to be repeated when the target data is changed. This leads to wastage of time spent for training. In future, the authors will be trying to evolve the objectives so that the inter-subject EEG classification can be addressed in a subject-invariant manner in order to eliminate the repeated training phase and thereby, broadening the application domain of the proposed method. 

\bibliographystyle{elsart-num}

\end{document}